\documentclass[showkeys,twocolumn,showpacs,preprintnumbers,amsmath,amssymb,prl]{revtex4}
\usepackage{Depken,graphicx}
\begin{document}
\title{Exact joint density-current probability function for the asymmetric exclusion process}
\author{Martin Depken and Robin Stinchcombe}
 \affiliation{University of Oxford, Department of Physics, Theoretical Physics\\ 1 Keble Road, Oxford, OX1 3NP, U.K.}
 \altaffiliation[MD's present address: ]{Instituut-Lorentz, Leiden University, P.O. Box 9506, 2300 RA Leiden, Netherlands\\ email: depken@lorentz.leidenuniv.nl}
\begin{abstract}
We study the asymmetric exclusion process with open boundaries and derive the exact form of the joint probability function for the occupation number and the current through the system. We further consider the thermodynamic limit, showing that the resulting distribution is non-Gaussian and that the density fluctuations have a discontinuity at the continuous phase transition, while the current fluctuations are continuous. The derivations are performed by using the standard operator algebraic approach, and by the introduction of new operators satisfying a modified version of the original algebra.
\end{abstract}


\maketitle

\begin{figure}[htp]
\hspace{-1cm}\includegraphics[width=.2\textwidth]{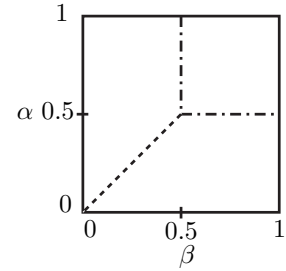}
\caption{\label{fig:pd} Phase diagram of the one dimensional exclusion process. The dashed line indicates the first order transition line, while the dash-dotted lines indicates the continuous transition lines.}
\end{figure}

Due to the lack of a general theory of non-equilibrium steady-states, a lot of the research in this area focuses on the study of simple models. Of special interest are stochastic interacting particle models~\cite{Liggett85,Liggett02}, and one hopes that many of the interesting qualitative characteristics of these simplified models are generic for a larger class of systems. A system that has received a lot of the attention is the partially asymmetric exclusion process (PASEP). Firstly this model is non-trivial, displaying steady-state phase transitions, yet simple enough to be integrable~\cite{Derrida92,Derrida93,Sasamoto99,Blythe99}, and further it maps onto certain growth models~\cite{Halpin-Healy95}, it models traffic flow~\cite{Schreckenberg98}, and it is believed to describe the large scale dynamics of the noisy burgers equation~\cite{Burgers74,Gwa92,Stinchcombe01} and the KPZ equation~\cite{Kardar86,Stinchcombe01}. There has been much progress in the analytical treatment of the PASEP, giving rise to host of exact results describing it's steady state properties~\cite{Derrida92,Derrida93,Sasamoto99,Blythe99,Ferrari94,Lee99,Derrida98,Derrida02}. At the heart of the distinction between non-equilibrium and equilibrium systems lies the ability of non-equilibrium systems to carry currents. So far, the results concerning the currents in different special cases of the PASEP~\cite{Ferrari94,Lee99,Praehofer02,Derrida03} are mainly for systems with periodic boundaries or infinite geometries with special initial conditions. In this letter we consider a finite system with open boundaries, but specialize the treatment to the one dimensional asymmetric exclusion process (ASEP). It consists of a lattice of size $L$, with site label $l$ running from left to right. Every site on the lattice can be occupied by no more than one particle. Given that the right neighboring site of an occupied site is empty, the occupying particle will jump to the empty site with a rate $1$. If the first site  on the lattice is unoccupied, particles are injected at this boundary with rate $\alpha$. Further given that we have a particle at the last site  of the lattice, it is ejected with the probability rate $\beta$. No further transitions are allowed. We will here limit our considerations to the case where we can view the boundary rates as deriving from particle reservoirs. We therefore take $0<\alpha=\rho_{\rm left}<1$ and $0<\beta=1-\rho_{\rm right}<1$, where $\rho_{\rm left}$ and $\rho_{\rm right}$ are the particle densities of the reservoirs. This model has been exactly solved~\cite{Derrida93} (see Figure~\ref{fig:pd} for the phase diagram) in the sense that the steady-state probability of any given microscopic configuration can (in principle) be calculated by applying a given set of algebraic rules. Even so, these calculations quickly become very cumbersome as the system size is increased.
Thus we here wish to extract general information about the system directly from the algebraic rules, without explicitly calculating the microscopic weights. Since the algebraic rules are instrumental to our later development we here give a very brief recap on their definition. The starting point is to represent any microscopic configurations in terms of a string of non-commuting operators $\opa D$ and $\opa E$, corresponding to a particle and a hole respectively. It can then be shown that the steady-state probability distribution can be written in terms of this operator string and two vectors, $\lv \alpha |$ and $|\beta\rv$, according to
\be{eq:ssm}
  P_{\rm ss}(\{n_l\})= (Z_L^{\alpha\beta})^{-1}\lv \alpha |\opa X(n_1)\opa X(n_2)\cdots\opa X(n_L)|\beta\rv.
\ee
Here the operator $\opa X(n_l)$ equals $\opa D$ if there is a particle at site $l$ ($n_l=1$), and $\opa E$ if site $l$ is unoccupied ($n_l=0$). The state independent factor  $Z_L^{\alpha\beta}=\lv \alpha |\opa (\bm D+\bm E)^L|\beta\rv$ ensures the proper normalization. For~(\ref{eq:ssm}) to hold true, the operators and  vectors must further satisfy the algebraic rules
\be{eq:opalg}
  \opa C\Def\opa D\opa E=\opa E+\opa D, \,\,\,\, \lv \alpha |\opa E=\frac{1}{\alpha}\lv \alpha |, \,\,\,\, \opa D |\beta\rv=\frac{1}{\beta}|\beta\rv,
\ee
where we have implicitly assumed that the normalizations of the vectors $|\alpha\rv$ and $\lv \beta|$ are such that $\lv \alpha |\beta\rv=1$. The algebraic rules~(\ref{eq:opalg}) are now all that is needed to calculate $P^{\alpha \beta}_{\rm ss}(\{n_l\})$, resulting in a polynomial of degree $L$ in $1/\alpha$ and $1/\beta$.

\begin{figure}[htp]
  \includegraphics[width=.45\textwidth]{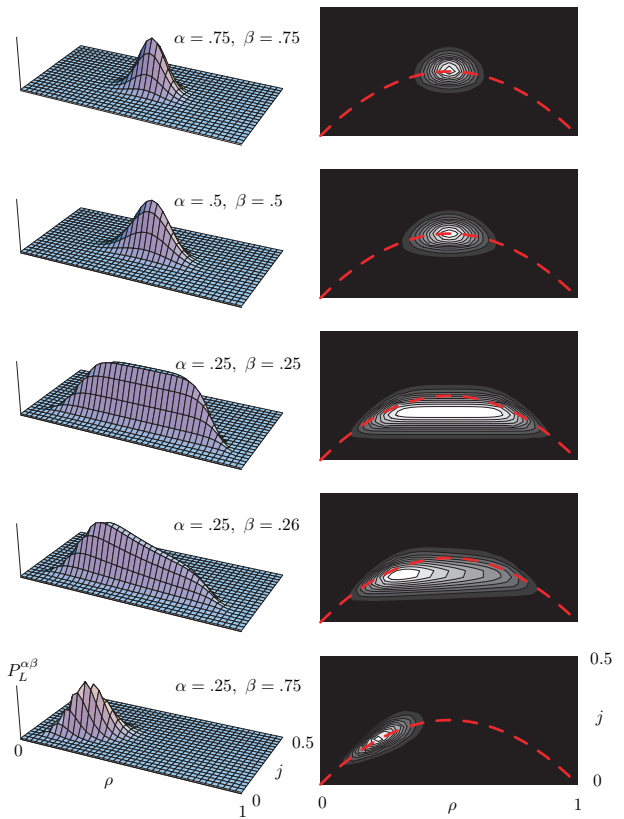}
\caption{\label{fig:Exact} Each row contains a surface and a contour plot of the exact probability distribution for the values of $\alpha$ and $\beta$ indicated, and with $\rho=N/L$ and $j=J/L$. The first three rows illustrate the behavior of the probability distribution as the system goes along the line of $\alpha=\beta$ through the critical point at $\alpha=\beta=.5$, while the last three graphs illustrate the behavior as the system goes through the first order transition at $\alpha=\beta=0.25$.  Overlaid in the contour plots (dashed line) is the curve $j=\rho(1-\rho)$ which defines the set of possible asymptotic average values of $\rho$ and $j$ throughout the systems different phases (not at the first order transition line). The system size is $L=40$.}
\end{figure}

In moving from a microscopic to a macroscopic view of the system we will here concentrate on the average density and current throughout the bulk. We derive the exact joint probability function for the average bulk current and density for any system size.  First we define the total activity within the system as the number of bulk bonds that can facilitate a transition of a particle in the immediate future, i.e. the total effective bulk transition rate. The bulk current is then defined as the activity divided by the system size. For any given state the activity equals the number of pairs of neighboring sites that has a particle to the left and a hole to the right. To get a handle on the activity, $J$, of a microscopic configuration of $N$ particles we choose to represent such a configuration by a sequence of $J$ objects of the form $\bm D^{p_j}\bm E^{h_j}$, possibly padded with $\bm E$'s to the left and $\bm D$'s to the right. Doing this we can write any microscopic steady state measure as
\begin{multline*}
P_{\rm ss}(\{n_l\})=\\ (Z_L^{\alpha\beta})^{-1}\lv \alpha |\opa E^{h_0} (\bm D^{p_1}\bm E^{h_1})\cdots (\bm D^{p_J}\bm E^{h_J})\bm D^{p_0}|\beta\rv,
\end{multline*}
by appropriately choosing the numbers $\{p_j,h_j\}$ and $J$. It further follows that the above expression is unique if $h_0,p_0\ge 0$ and the rest satisfy $h_j,p_j\ge 1$. We can now in principle calculate the joint probability distribution for $N$ and $J$ by summing the above over all $h_j$'s and $p_j$'s consistent with a specific number of particles ($\sum_{j=0}^J p_j=N$) and a given system size ($N+\sum_{j=0}^J h_j=L$). Choosing to enforce these constraints with contour-integral representations of the Kronecker delta, the expression for the joint particle-activity probability function can be written as
\begin{multline}
\label{eq:obsprob}
  P^{\alpha\beta}_L(N,J)=\frac{\alpha\beta}{Z_L}\oint\limits_{C_{z},C_{\bar z}} \frac{\d z\d \bar z}{(2\pi\imath)^2}\frac{1}{z^{N+1}\bar z^{L-N+1}}\\\frac{1}{(z-\beta)(\bar z-\alpha)}\lv \alpha |\lp\sum\limits_{p=1}^{N} (z \bm D)^p \sum\limits_{h=1}^{L-N}(\bar z \bm E)^h\rp^J|\beta\rv.
\end{multline}
Here $C_z$ ($C_{\bar z}$) is a directed contour that encircle the pole at the origin of the complex $z$ ($\bar z$) plane once in the positive direction, with $|z|<\beta$ ($|\bar z|<\alpha$). The first step toward explicitly calculating~(\ref{eq:obsprob}) is through considering the properties of the operators $\sum_p (z\bm D)^p$ and $\sum_h (\bar z\bm E)^h$. Surprisingly it turns out~\cite{Depken03} that a slight modification of the above operators
$$
\begin{array}{rl}
  \bm D'&\Def (1-(z+\bar z))\bm D \sum\limits_{h=0}^{N-1} (z\bm D)^h,\\
\bm E'&\Def(1-(z+\bar z))\bm E\sum\limits_{p=0}^{L-N-1} (\bar z\bm E)^p,
\end{array}
$$
satisfy the ``relaxed'' operator algebra
$$
  \bm D'\bm E'=\bm D'+\bm E'+\O(z^{N},\bar z^{L-N}).
$$
The relaxed eigenvectors and eigenvalues are further given by
$$
  \bm D'|\beta\rv =|\beta\rv\frac{1}{\beta'}+\O(z^N), \quad  \lv \alpha |\bm E'=\frac{1}{\alpha'}\lv \alpha |+\O(\bar z^{L-N}),
$$
with
$$
\alpha'\Def\frac{\alpha-\bar z}{1-(z+\bar z)}, \quad
\beta'\Def\frac{\beta-z}{1-(z+\bar z)}.
$$
The fact that these eigenvalues are complex is of no concern since we consider only finite polynomials in the inverse eigenvalues. Any result is thus uniquely extendable into the complex plane through analytic continuation. We can now rewrite~(\ref{eq:obsprob}) in terms of the primed operators, and start using the relaxed operator algebra to transform the expression.  The result of any such manipulation would be the same, up to terms of order $z^{N}$ and $\bar z^{L-N}$, as if the operator algebra would have been exact. The extra terms have no effect under the contour integral in~(\ref{eq:obsprob})  since the poles at the origins are both of order equal or lower than $N$ and $L-N$. (The case for $J=0$ is trivial.) Thus, using the relaxed algebra to perform any manipulation within~(\ref{eq:obsprob}) is equivalent to using an exact algebra. Therefore we can write
\begin{multline}\label{eq:prob1}
   P^{\alpha\beta}_L(N,J)=\frac{\alpha\beta}{Z_L^{\alpha\beta}}\oint\limits_{C_{z},C_{\bar z}} \frac{\d z\d \bar z}{(2\pi\imath)^2}\frac{1}{z^{N+1-J}\bar z^{L-N+1-J}}\\ \frac{Z_J^{\alpha'\beta'}}{(z-\beta)(\bar z-\alpha)(1-(z+\bar z))^{2J}}.
\end{multline}
This expression is the main result of this letter, and since all quantities in it are known exactly, it yields both the exact finite system size form of $P^{\alpha\beta}_L(N,J)$, as well as the asymptotic form in the large system size limit. By first calculating the generating functional $G(\mu)=\sum_{\mu=0}^{\infty}\mu^L Z^{\alpha\beta}_L$~\cite{Depken03} (which has recently been derived in a different manner in~\cite{Blythe04}), the above further yields a double contour integral expression for the generating functional of $N$ and $J$~\cite{Depken03}. Below we present exact and asymptotic results for $P_L^{\alpha\beta}(N,J)$.
\paragraph{Finite systems:} The integral in~(\ref{eq:prob1}) is easily calculated with the help Cauchy's integral theorem. All we need to do is to calculate the coefficient of the term proportional to $(z\bar z)^{-1}$ in the Laurent-series expansion of the integrand in~(\ref{eq:prob1}). For $J=0$  we have $Z_J^{\alpha'\beta'}=1$ and thus
$$
  P^{\alpha\beta}_L(N,0)=\frac{1}{Z_L^{\alpha\beta}}\lp1/\beta\rp^N\lp1/\alpha\rp^{L-N},
$$
which is obviously correct since the inactive state must have $L-N$ empty sites followed by $N$ filled sites. For $J\ge 1$ we use the exact form of the normalization function~\cite{Derrida93},
$$
  Z^{\alpha\beta}_L=\sum\limits_{l=1}^{L} A_{L,l}\sum\limits_{k=0}^{l}\frac{1}{\alpha^k\beta^{l-k}},\quad A_{L,l}=\frac{l(2L-l-1)!}{L!(L-l)!}
$$
to write 
\begin{multline}
\label{eq:K}
  P^{\alpha\beta}_L(N,J)=\frac{\alpha\beta}{Z_L^{\alpha\beta}}\sum\limits_{j=1}^J A_{J,j} \sum\limits_{k=0}^j \sum\limits_{c=0}^{L-N-J}\sum\limits_{d=0}^{N-J}\\ G_{k,c}(\alpha)G_{j-k,d}(\beta)H_{2J-j,L-N-J-c,N-J-d},
\end{multline}
with
\begin{eqnarray*}
  G_{k,c}^{\alpha}&=&\bin{k+c}{c}\frac{1}{\alpha^{c+k+1}}\\
  H_{K,d,e}&=&\bin{K-1+d+e}{d+e}\bin{d+e}{e}.
\end{eqnarray*}
Through the above we now have the exact form of the sought-after joint probability function for any system size. The form is illustrated in Figure~\ref{fig:Exact}.

\paragraph{The thermodynamics limit:}
We here return to (\ref{eq:prob1}). Using the asymptotic form of the normalizing function given in~\cite{Derrida93}, we perform a steepest descent calculation to get the asymptotic results. We consider the different phases individually. Due to the particle-hole symmetry $P_L^{\alpha\beta}(N,J)=P_L^{\beta\, \alpha}(L-N,J)$ it is only necessary to explicitly consider the case $\alpha<\beta$.

\begin{figure}[htp]
  \includegraphics[width=.45\textwidth]{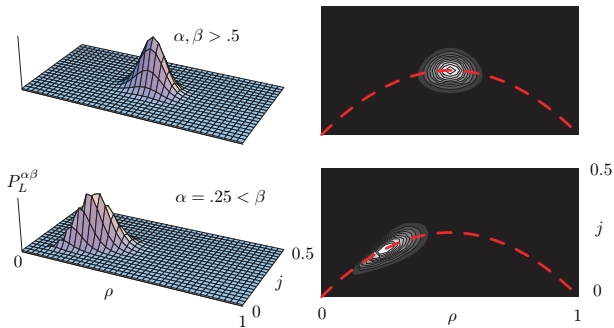}\caption{\label{fig:Asympt} The two rows display a surface and a contour plot of the leading behavior of the asymptotic probability distribution. The calculations were performed at the injection and ejection rates indicated and at a system of size $L=40$ (to make the result comparable to Figure~\ref{fig:Exact}).}
\end{figure}

First turning to the maximal-current phase we consider~(\ref{eq:prob1}) and drop all pre-factors that are independent of $N$ and $J$ (this will be done throughout) to write
\begin{multline*}
  P^{\alpha\beta}_L(N,J)\sim \frac{4^{J}}{J^{3/2}}\oint\limits_{C_{z},C_{\bar z}} \frac{\d z\d \bar z}{(2\pi\imath)^2}\frac{1}{z^{N-J+1}\bar z^{L-N-J+1}}\\\frac{1}{(1-(z+\bar z))^{2J-1}}\frac{1}{(2\alpha-1+z-\bar z)^2(2\beta-1-(z-\bar z))^2}.
\end{multline*}
The asymptotic behavior of these integrals is in principle straight forward to calculate. In practice though, it turns out to be quite cumbersome since one has to determine which of the saddle points and lower order poles give the dominant contributions. We can shortcut this through only considering the asymptotic form in some finite region around the peak of the distribution. From~\cite{Derrida93} we know that the average density and current is $\alpha$ and $\beta$ independent. Thus, the lower order poles cannot dictate the asymptotic behavior around the peak value of the probability distribution, and instead this must be set by the saddle points
$$
  z^*=\rho-j, \quad \bar z^*=1-\rho-j, \quad \rho= N/L,\quad j= J/L. 
$$
A saddle-point approximation thus results in
\begin{multline}\label{eq:hc}
P^{\alpha\beta}_L(\rho,j)\sim 
\lp\frac{1}{j^{2j}(\rho-j)^{\rho-j}(1-\rho-j)^{1-\rho-j}}\rp^L,
\end{multline}
where we for simplicity have dropped all the sub-dominant pre-factors. Even though the extent of the region of validity of~(\ref{eq:hc}) is unknown, it should be pointed out that the size of this region is finite (as long as the system is away from any phase boundaries) and independent of system size. In the first row of Figure~\ref{fig:Asympt} we show the resulting dominating asymptotic plots.

Now turning to the low-current phases we have
\begin{multline*}
 P^{\alpha\beta}_L(\rho,j)\sim\oint\limits_{C_{z},C_{\bar z}} \frac{\d z\d \bar z}{(2\pi\imath)^2}\frac{1}{z^{N-J+1}\bar z^{L-N-J+1}}\\
\frac{2\alpha-1+z-\bar z}{\beta-\alpha-(z-\bar z)}\frac{1}{(\alpha-\bar z)^{J+1}(1-z-\alpha)^{J+1}}.
\end{multline*}
The same arguments as applied in the maximal-current phase now gives us the asymptotic probability distribution around the peak. Again it is the saddle points
$$
z^*=\frac{\rho-j}{\rho}(1-\alpha), \quad \bar z^*=\frac{1-\rho-j}{1-\rho}\alpha
$$
that dominate. The resulting dominant form is
\begin{multline}
\label{eq:lc}
  P^{\alpha\beta}_L(N,J)\sim
\\
\lp\frac{\rho^\rho (1-\rho)^{1-\rho}}{\alpha^{1-\rho}(1-\alpha)^{\rho}}\frac{1}{j^{2j}(\rho-j)^{\rho-j}(1-\rho-j)^{1-\rho-j}}\rp^L.
\end{multline}
The above result, valid for the low-density phase, is directly transferable to the high-density phase through the use of the particle hole symmetry mentioned above.  A realization of the asymptotically dominating part in the low-density phase is shown in the second row of~Figure~\ref{fig:Asympt}. 

It is clear from the asymptotic forms that the probability distribution is non-Gaussian in all phases. This is consistent with the view that long-range correlations are a generic feature of non-equilibrium systems~\cite{Katz83}.  In general, as a phase-transition line is approached the border of the region of validity of the asymptotic forms~(\ref{eq:hc}) and (\ref{eq:lc}) must approach the peak. It should be further pointed out that as the continuum transition is passed, the asymptotic forms~(\ref{eq:hc}) and~(\ref{eq:lc}) indicates that there will be a finite jump in the connected density-density correlator. Comparing this to equilibrium systems, these transitions correspond to proper second order transitions. 

In conclusion we note that it would be very interesting to examine if the same ``trick'' of introducing a relaxed algebra could somehow be applied to the PASEP. This especially since this model interpolates between a equilibrium and non-equilibrium steady state. It would further be interesting to derive the full asymptotic form of the probability distribution since sufficiently close to a phase transition, any finite system will reach a point at which the region of validity of the above asymptotic forms shrink to the size of the typical fluctuations. In this region the system crosses over to a situation where the asymptotic fluctuations are governed by the tails excluded in the above development.

This work was supported by EPSRC under the Oxford Condensed Matter Theory Grant No. GR/R83712/01 and GR/M04426. MD gratefully acknowledges support from the Merton College Oxford Domus scholarship fund, and the Royal Swedish Academy of Science.  
\bibliography{bibliography}

\begin{thebibliography}{21}
\expandafter\ifx\csname natexlab\endcsname\relax\def\natexlab#1{#1}\fi
\expandafter\ifx\csname bibnamefont\endcsname\relax
  \def\bibnamefont#1{#1}\fi
\expandafter\ifx\csname bibfnamefont\endcsname\relax
  \def\bibfnamefont#1{#1}\fi
\expandafter\ifx\csname citenamefont\endcsname\relax
  \def\citenamefont#1{#1}\fi
\expandafter\ifx\csname url\endcsname\relax
  \def\url#1{\texttt{#1}}\fi
\expandafter\ifx\csname urlprefix\endcsname\relax\def\urlprefix{URL }\fi
\providecommand{\bibinfo}[2]{#2}
\providecommand{\eprint}[2][]{\url{#2}}

\bibitem[{\citenamefont{Liggett}(1985)}]{Liggett85}
\bibinfo{author}{\bibfnamefont{T.~M.} \bibnamefont{Liggett}},
  \emph{\bibinfo{title}{Interacting particle systems}}
  (\bibinfo{publisher}{Springer-Verlag, New York}, \bibinfo{year}{1985}).

\bibitem[{\citenamefont{Liggett}(2002)}]{Liggett02}
\bibinfo{author}{\bibfnamefont{T.~M.} \bibnamefont{Liggett}}
  (\bibinfo{year}{2002}), \bibinfo{note}{lectures from the school and
  conference on Probability theory, Trieste, Italy}.

\bibitem[{\citenamefont{Derrida et~al.}(1992)\citenamefont{Derrida, Domany, and
  Mukamel}}]{Derrida92}
\bibinfo{author}{\bibfnamefont{B.}~\bibnamefont{Derrida}},
  \bibinfo{author}{\bibfnamefont{E.}~\bibnamefont{Domany}}, \bibnamefont{and}
  \bibinfo{author}{\bibfnamefont{D.}~\bibnamefont{Mukamel}},
  \bibinfo{journal}{J. Stat. Phys.} \textbf{\bibinfo{volume}{69}},
  \bibinfo{pages}{667} (\bibinfo{year}{1992}).

\bibitem[{\citenamefont{Derrida et~al.}(1993)\citenamefont{Derrida, Evans,
  Hakim, and Pasquier}}]{Derrida93}
\bibinfo{author}{\bibfnamefont{B.}~\bibnamefont{Derrida}},
  \bibinfo{author}{\bibfnamefont{M.}~\bibnamefont{Evans}},
  \bibinfo{author}{\bibfnamefont{V.}~\bibnamefont{Hakim}}, \bibnamefont{and}
  \bibinfo{author}{\bibfnamefont{V.}~\bibnamefont{Pasquier}},
  \bibinfo{journal}{J. Phys. A} \textbf{\bibinfo{volume}{26}},
  \bibinfo{pages}{1493} (\bibinfo{year}{1993}).

\bibitem[{\citenamefont{Sasamoto}(1999)}]{Sasamoto99}
\bibinfo{author}{\bibfnamefont{T.}~\bibnamefont{Sasamoto}},
  \bibinfo{journal}{J. Phys. A} \textbf{\bibinfo{volume}{32}},
  \bibinfo{pages}{7109} (\bibinfo{year}{1999}).

\bibitem[{\citenamefont{Blythe et~al.}(1999)\citenamefont{Blythe, Evans,
  Colaiori, and Essler}}]{Blythe99}
\bibinfo{author}{\bibfnamefont{R.}~\bibnamefont{Blythe}},
  \bibinfo{author}{\bibfnamefont{M.}~\bibnamefont{Evans}},
  \bibinfo{author}{\bibfnamefont{F.}~\bibnamefont{Colaiori}}, \bibnamefont{and}
  \bibinfo{author}{\bibfnamefont{F.}~\bibnamefont{Essler}},
  \bibinfo{journal}{J. Phys. A} \textbf{\bibinfo{volume}{33}},
  \bibinfo{pages}{2313} (\bibinfo{year}{1999}).

\bibitem[{\citenamefont{Halpin-Healy and Zhang}(1995)}]{Halpin-Healy95}
\bibinfo{author}{\bibfnamefont{T.}~\bibnamefont{Halpin-Healy}}
  \bibnamefont{and} \bibinfo{author}{\bibfnamefont{Y.-C.} \bibnamefont{Zhang}},
  \bibinfo{journal}{Phys. Rep.} \textbf{\bibinfo{volume}{254}},
  \bibinfo{pages}{215} (\bibinfo{year}{1995}).

\bibitem[{\citenamefont{Schreckenberg and Wolf}(1998)}]{Schreckenberg98}
\bibinfo{editor}{\bibfnamefont{M.}~\bibnamefont{Schreckenberg}}
  \bibnamefont{and} \bibinfo{editor}{\bibfnamefont{D.}~\bibnamefont{Wolf}},
  eds., \emph{\bibinfo{title}{Traffic and Granular Flow '97}}
  (\bibinfo{publisher}{Springer-Verlag, Singapore}, \bibinfo{year}{1998}).

\bibitem[{\citenamefont{Burgers}(1974)}]{Burgers74}
\bibinfo{author}{\bibfnamefont{J.}~\bibnamefont{Burgers}},
  \emph{\bibinfo{title}{The non-linear diffusion equation}}
  (\bibinfo{publisher}{Boston MA, Reidel}, \bibinfo{year}{1974}).

\bibitem[{\citenamefont{Gwa and Spohn}(1992)}]{Gwa92}
\bibinfo{author}{\bibfnamefont{L.}~\bibnamefont{Gwa}} \bibnamefont{and}
  \bibinfo{author}{\bibfnamefont{H.}~\bibnamefont{Spohn}},
  \bibinfo{journal}{Phys. Rev. A} \textbf{\bibinfo{volume}{46}},
  \bibinfo{pages}{844} (\bibinfo{year}{1992}).

\bibitem[{\citenamefont{Stinchcombe}(2001)}]{Stinchcombe01}
\bibinfo{author}{\bibfnamefont{R.}~\bibnamefont{Stinchcombe}},
  \bibinfo{journal}{Adv. Phys.} \textbf{\bibinfo{volume}{50}},
  \bibinfo{pages}{431} (\bibinfo{year}{2001}).

\bibitem[{\citenamefont{Kardar et~al.}(1986)\citenamefont{Kardar, Parisi, and
  Zhang}}]{Kardar86}
\bibinfo{author}{\bibfnamefont{M.}~\bibnamefont{Kardar}},
  \bibinfo{author}{\bibfnamefont{G.}~\bibnamefont{Parisi}}, \bibnamefont{and}
  \bibinfo{author}{\bibfnamefont{Y.-C.} \bibnamefont{Zhang}},
  \bibinfo{journal}{Phys. Rev. Lett.} \textbf{\bibinfo{volume}{56}},
  \bibinfo{pages}{889} (\bibinfo{year}{1986}).

\bibitem[{\citenamefont{Ferrari and Fontes}(1994)}]{Ferrari94}
\bibinfo{author}{\bibfnamefont{P.~A.} \bibnamefont{Ferrari}} \bibnamefont{and}
  \bibinfo{author}{\bibfnamefont{L.~R.~G.} \bibnamefont{Fontes}},
  \bibinfo{journal}{Ann. Prob.} \textbf{\bibinfo{volume}{22}},
  \bibinfo{pages}{820} (\bibinfo{year}{1994}).

\bibitem[{\citenamefont{Lee and Kim}(1999)}]{Lee99}
\bibinfo{author}{\bibfnamefont{D.-S.} \bibnamefont{Lee}} \bibnamefont{and}
  \bibinfo{author}{\bibfnamefont{D.}~\bibnamefont{Kim}},
  \bibinfo{journal}{Phys. Rev. E} \textbf{\bibinfo{volume}{59}},
  \bibinfo{pages}{6476} (\bibinfo{year}{1999}).

\bibitem[{\citenamefont{Derrida}(1998)}]{Derrida98}
\bibinfo{author}{\bibfnamefont{B.}~\bibnamefont{Derrida}},
  \bibinfo{journal}{Phys. Rep.} \textbf{\bibinfo{volume}{301}},
  \bibinfo{pages}{65} (\bibinfo{year}{1998}).

\bibitem[{\citenamefont{Derrida et~al.}(2002)\citenamefont{Derrida, Lebowitz,
  and Speer}}]{Derrida02}
\bibinfo{author}{\bibfnamefont{B.}~\bibnamefont{Derrida}},
  \bibinfo{author}{\bibfnamefont{J.}~\bibnamefont{Lebowitz}}, \bibnamefont{and}
  \bibinfo{author}{\bibfnamefont{E.}~\bibnamefont{Speer}},
  \bibinfo{journal}{Phys. Rev. Lett.} \textbf{\bibinfo{volume}{89}},
  \bibinfo{pages}{030601} (\bibinfo{year}{2002}).

\bibitem[{\citenamefont{Praehofer and Spohn}(2002)}]{Praehofer02}
\bibinfo{author}{\bibfnamefont{M.}~\bibnamefont{Praehofer}} \bibnamefont{and}
  \bibinfo{author}{\bibfnamefont{H.}~\bibnamefont{Spohn}},
  \bibinfo{journal}{Prog. Prob.} \textbf{\bibinfo{volume}{51}},
  \bibinfo{pages}{185} (\bibinfo{year}{2002}).

\bibitem[{\citenamefont{Derrida et~al.}(2003)\citenamefont{Derrida, Dou\c{c}ot,
  and Roche}}]{Derrida03}
\bibinfo{author}{\bibfnamefont{B.}~\bibnamefont{Derrida}},
  \bibinfo{author}{\bibfnamefont{B.}~\bibnamefont{Dou\c{c}ot}},
  \bibnamefont{and} \bibinfo{author}{\bibfnamefont{P.-E.} \bibnamefont{Roche}},
  \bibinfo{journal}{cond-mat/0310453}  (\bibinfo{year}{2003}).

\bibitem[{\citenamefont{Depken}(2003)}]{Depken03}
\bibinfo{author}{\bibfnamefont{M.}~\bibnamefont{Depken}}, Ph.D. thesis,
  \bibinfo{school}{Department of Physics, University of Oxford}
  (\bibinfo{year}{2003}), \bibinfo{note}{{M}. Depken and {R}. Stinchcombe, {\it
  to be published}}.

\bibitem[{\citenamefont{Blythe et~al.}(2004)\citenamefont{Blythe, Janke,
  Johnston, and Kenna}}]{Blythe04}
\bibinfo{author}{\bibfnamefont{R.}~\bibnamefont{Blythe}},
  \bibinfo{author}{\bibfnamefont{W.}~\bibnamefont{Janke}},
  \bibinfo{author}{\bibfnamefont{D.}~\bibnamefont{Johnston}}, \bibnamefont{and}
  \bibinfo{author}{\bibfnamefont{R.}~\bibnamefont{Kenna}},
  \bibinfo{journal}{cond-mat/0401385}  (\bibinfo{year}{2004}).

\bibitem[{\citenamefont{Katz et~al.}(1983)\citenamefont{Katz, Lebowitz, and
  Spohn}}]{Katz83}
\bibinfo{author}{\bibfnamefont{S.}~\bibnamefont{Katz}},
  \bibinfo{author}{\bibfnamefont{J.}~\bibnamefont{Lebowitz}}, \bibnamefont{and}
  \bibinfo{author}{\bibfnamefont{H.}~\bibnamefont{Spohn}}, \bibinfo{journal}{J.
  Rev. B} \textbf{\bibinfo{volume}{28}}, \bibinfo{pages}{1655}
  (\bibinfo{year}{1983}).

\end{thebibliography}

\end{document}